\def\be{\begin{equation}} 
\def\ee{\end{equation}} 
\def\bea{\begin{eqnarray*}} 
\def\eea{\end{eqnarray*}}

\documentclass[useAMS,usenatbib]{mn2e} 
\usepackage{subfigure} 
\usepackage{graphicx} 
\usepackage{flafter} 
\usepackage{lscape} 
\begin{document} 
 
\title{A bright, dust-obscured, millimeter-selected galaxy beyond the  
Bullet Cluster (1E0657-56)} 
 
 
\author[G.W. Wilson et al.]{G.W.~Wilson$^1$, 
D.H.~Hughes$^2$, 
I.~Aretxaga$^2$ , 
H.~Ezawa$^3$,  
J.E.~Austermann$^1$, 
\newauthor 
S.~Doyle$^4$, 
D.~Ferrusca$^{5,2}$,
I.~Hern\'andez-Curiel$^{2,6}$, 
R.~Kawabe$^3$, 
T.~Kitayama$^7$, 
K.~Kohno$^8$, 
\newauthor 
A.~Kuboi$^3$, 
H.~Matsuo$^3$, 
P.D.~Mauskopf$^4$, 
Y.~Murakoshi$^7$, 
A.~Monta\~{n}a$^2$, 
\newauthor  
P.~Natarajan$^9$, 
T.~Oshima$^3$, 
N.~Ota$^{10,11}$, 
T.A.~Perera$^1$, 
J.~Rand$^1$ , 
K.S.~Scott$^1$, 
\newauthor 
K.~Tanaka$^8$ , 
M.~Tsuboi$^{10}$, 
C.C.~Williams$^1$, 
N.~Yamaguchi$^3$,
M.S.~Yun$^1$\\ 
$^1$Department of Astronomy, University of Massachusetts, Amherst, MA 01003, USA.\\ 
$^2$Instituto Nacional de Astrof\'isica, \'Optica y Electr\'onica, Tonantzintla, 
Aptdo. Postal 51 y 216, 72000 Puebla, Pue., Mexico.\\ 
$^3$National Astronomical Observatory of Japan, 2-21-1 Osawa, Mitaka, Tokyo 181-8588, Japan.\\ 
$^4$Department of Physics and Astronomy, Cardiff University, Cardiff CF24 3YB, Wales, UK.\\ 
$^5$Department of Physics, University of California, Berkeley, 94720-7300, USA. \\
$^6$Centro de Radioastronom\'{\i}a y Astrof\'{\i}sica, Universidad Nacional  
Aut\'onoma de M\'exico, Aptdo. Postal 72-3 (Xangari), 58089 Morelia, Mexico.\\ 
$^7$Department of Physics, Toho University, Funabashi, Chiba 274-8510, Japan\\ 
$^8$Institute of Astronomy, The University of Tokyo, 2-21-1 Osawa, Mitaka, Tokyo, 181-0015, Japan.\\ 
$^9$Department of Physics, Yale University, PO Box 208120, New Haven, CT 06520-208120, USA.\\ 
$^{10}$Institute of Space and Astronautical Science, JAXA, 3-1-1 Yoshinodai, Sagamihara, Kanagawa, 229-8510, Japan.\\ 
$^{11}$Max-Planck-Intitut f\"{u}r  extraterrestrische Physik, Giessenbachstrasse, 85748 Garching, Germany 
} 
 
\date{\today} 
 
\pagerange{\pageref{firstpage}--\pageref{lastpage}} \pubyear{2007} 
 
\maketitle 
 
\label{firstpage} 
 
\begin{abstract} 
Deep 1.1~mm continuum observations of 1E0657-56 (the ``Bullet
Cluster'') taken with the millimeter-wavelength camera AzTEC on the
10-m Atacama Submillimeter Telescope Experiment (ASTE), have revealed
an extremely bright (S$_{\rm{1.1mm}}=15.9$~mJy) unresolved source.
This source, MMJ065837-5557.0, lies close to a maximum in the density
of underlying mass-distribution, towards the larger of the two
interacting clusters as traced by the weak-lensing analysis of
\citet{clowe2006}.  Using optical--IR colours we argue that
MMJ065837-5557.0 lies at a redshift of $z = 2.7 \pm 0.2$. A
lensing-derived mass-model for the Bullet Cluster shows a
critical-line (caustic) of magnification within a few arcsecs of the
AzTEC source, sufficient to amplify the intrinsic
millimeter-wavelength flux of the AzTEC galaxy by a factor of $\gg
20$.  After subtraction of the foreground cluster emission at 1.1mm
due to the Sunyaev-Zel'dovich effect, and correcting for the
magnification, the rest-frame FIR luminosity of MMJ065837-5557.0 is
$\le 10^{12} \rm L_{\odot}$, characteristic of a luminous infrared
galaxy (LIRG). We explore various scenarios to explain the colors,
morphologies and positional offsets between the potential optical and
IR counterparts, and their relationship with MMJ065837-5557.0.  Until
higher-resolution and more sensitive (sub)millimeter observations are
available, the detection of background galaxies close to the caustics
of massive lensing clusters offers the only opportunity to study this
intrinsically faint millimeter-galaxy population.
\end{abstract}


\section{INTRODUCTION} 
A high-redshift ($z \gg 1$) strongly-evolving submillimeter galaxy
population (hereafter SMGs) has been identified in a series of
continuum imaging surveys during the last decade. The most significant
surveys have been undertaken with SCUBA at 850$\mu$m
(\citet{coppin2006} and references therein), MAMBO at 1.2mm
\citep{greve2004,bertoldi2007}, BOLOCAM \citep{laurent2005} at 1.1~mm,  
and more recently with AzTEC at 1.1mm \citep{scott2008}.  These 
collective submillimeter surveys, together with comprehensive 
multi-wavelength imaging and spectroscopic follow-up spanning X-ray to 
radio wavelengths, have demonstrated that SMGs are representative of a 
population of massive, dust-enshrouded, optically-faint galaxies 
undergoing significant starformation with rates $\gg 200 \rm 
M_{\odot}/yr^{-1}$. The median redshift of the bright SMG population, 
based on spectroscopic and photometric redshifts, is $\sim 2.5$, with 
approximately 50\% of the population at $1.9 < z < 2.9$ 
\citep{chapman2003,chapman2005,aretxaga2003,aretxaga2007,pope2005}.  
Scaling the observed flux-densities of all SMGs with detections in the 
above flux-limited surveys to 850$\mu$m, the majority of the 
population have been detected with flux densities of $\rm 2\,mJy < 
S_{\rm{850}~\mu\rm{m}} < 12\,mJy$. 
 
Making the connection between ultra-luminous SMGs and their local
analogs - presumably ULIRGS and LIRGS - requires either larger
telescopes such as the Large Millimeter Telescope (LMT) and the
Atacama Large Millimeter Array (ALMA), or using the presence of
foreground structure to amplify the faint background galaxies.  To
date, the deepest SMG surveys have been conducted towards
lensing-clusters that amplify an intrinsically fainter population of
SMGs in the high-$z$ universe
\citep{smail1997,smail2002, chapman2002, cowie2002, kneib2004, knudsen2006}.   
Imaging massive clusters offers the advantages of amplification
of the background universe due to the gravitational-potential of the
cluster and, in general, an increased effective resolution in the
source plane (and correspondingly decreased survey confusion limit).
Unfortunately, this same advantage for those studying faint background
galaxies comes at a cost to those interested in studying the clusters
themselves using the Sunyaev-Zel'dovich Effect (SZE).  The amplified
(sub)millimeter-wavelength emission from members of the SMG field
population that are lensed by the large-scale cluster potential and by
the increased probability of galaxy-galaxy lensing along the line of
sight to the cluster, increase the likelihood of point-source
contamination of the SZE.
 
In this paper we present the discovery of an extremely bright
millimeter-selected SMG (MMJ065837-5557.0) in the direction of the
massive X-ray luminous Bullet Cluster (z=0.297).  The 1.1
mm-wavelength observations were made with the AzTEC instrument
\citep{wilson2008} on the Atacama Submillimeter Telescope Experiment 
(ASTE, \citet{ezawa2004}). These AzTEC data were undertaken as part of 
a larger project to trace the evolution in the surface-density of SMGs 
towards over-dense regions in the low and high-redshift 
Universe. Using archival HST and Spitzer data, we identify potential 
optical and IR counterparts to the AzTEC source. The 
photometric-colors, and the small positional-offset of the AzTEC 
source and the optical--IR counterparts from the critical 
magnification-line derived from cluster mass-models strongly support 
the suggestion that MMJ065837-5557.0 is a moderately-luminous high-$z$ 
galaxy that has been lensed and amplified by the massive foreground 
Bullet Cluster. 
 
Throughout this paper we adopt the following  
cosmological model: a Hubble constant $H_0 = 72 \rm ~kms^{-1} Mpc^{-1}$,  
and density parameters $\Omega_M = 0.3$ and $\Omega_\Lambda = 0.7$.

 
\section{AzTEC/ASTE observations} 
\label{sec:data} 
 
AzTEC is a 144-element bolometric receiver currently tuned to operate
in the 1.1mm atmospheric window~\citep{wilson2008}.  We have
previously used AzTEC to complete a successful set of observations of
the submillimeter galaxy population in blank-fields from the 15~m
diameter James Clerk Maxwell Telescope (e.g. \citet{scott2008,
perera2008}; Austermann et al, in preparation).  In 2007 AzTEC was
mounted on the ASTE telescope, a 10~m diameter antenna located on
Pampa La Bola, near Cerro Chajnantor in Chile. ASTE provides AzTEC
with an angular resolution of 30\arcsec\ FWHM. One season of
observations (June--October 2007) from this excellent high ($\sim
4900$m) and dry site in the Atacama Desert has been completed and
another season will begin in July 2008.
 
\begin{figure} 
\includegraphics[width=\hsize]{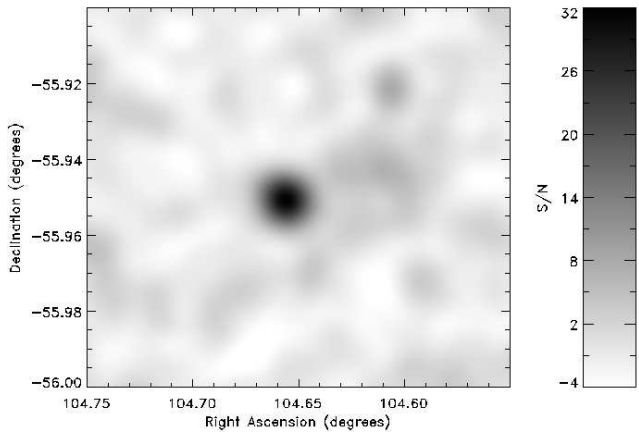} 
\caption{AzTEC image of MMJ065837-5557.0.  The colorbar shows the shading  
         in units of signal-to-noise (S/N) of a point source
         detection.  The underlying noise in the map varies by
         4.6\% across the portion of the image shown.  The faint
         diffuse emission to the west of the point source is
         dominated by the extended SZE signal from the Bullet cluster
         (Ezawa et al. in preparation) however due to the point
         source filtering of the map, no morphological information
         about the SZE should be taken from this figure.}
\label{fig:stupid_image}  
\end{figure}

The bright millimeter-wavelength source, shown in
Figure~\ref{fig:stupid_image}, was found in our 202 sq. arcmin AzTEC
imaging survey of the Bullet Cluster and its environment (Ezawa et
al., in preparation).  Observations were made by continuously scanning
the telescope boresight in azimuth and elevation in a modified
Lissajous pattern centered on 06h58m29.40s, -55d56m42.0s.  The
modified Lissajous pattern is defined as a function of time, $t$, by
\bea
\delta \mathrm{Az} &=& 5.5^{\prime}\sin{at} + 2^{\prime}\sin{at/30.} \\ 
\delta \mathrm{El} &=& 5.5^{\prime}\sin{bt} + 2^{\prime}\sin{bt/30} 
\eea 
where $a/b=8/9$ and $\delta$Az and $\delta$El are physical coordinates
relative to the field center.  The actual values of $a$ and $b$ are
normalized to limit the peak telescope velocity to 300\arcsec per
second.  A total of 61 maps of the cluster were completed, each taking
41.7 minutes, and collectively yielding an equivalent total on-source
integration time of 193~s per $3\arcsec\times3\arcsec$ pixel.  The
data are reduced and co-added using the standard AzTEC data analysis
pipeline with very similar techniques to those described in
\citet{scott2008}.  The resultant map has an average point-source
flux-error of 0.54~mJy over the inner 150~arcmin$^2$ area.  A full
analysis of the extended 1.1~mm Sunyaev-Zel'dovich emission of the
cluster and other point sources in the map will be presented in
subsequent papers.

\subsection{Pointing Corrections and Source Position} 
\label{sec:fit} 
A small correction to the telescope pointing model is applied to all
observations of the Bullet Cluster field based on periodic
observations of a bright point source, PKS 0537-441, with a 1.1~mm
flux of $\sim$5~Jy.  We derived the correction by linearly
interpolating the pointing offsets measured from
$4\arcmin\times4\arcmin$ maps of the point source observed every
2~hours.  Pointing measurements always bracketed observations of the
cluster.  Our resulting pointing uncertainty, as measured from
stacking the 1.1~mm flux at the locations of 212 known radio
sources in our map of the GOODS-S field (using the technique described
in \citet{scott2008}) is 4.8\arcsec\ rms.  This is a random pointing 
error which broadens the PSF of the instrument by less than 
0.5\arcsec and suppresses the point source response by an amount that 
is negligible compared to the photometric error. 
 
\begin{figure} 
\includegraphics[width=\hsize]{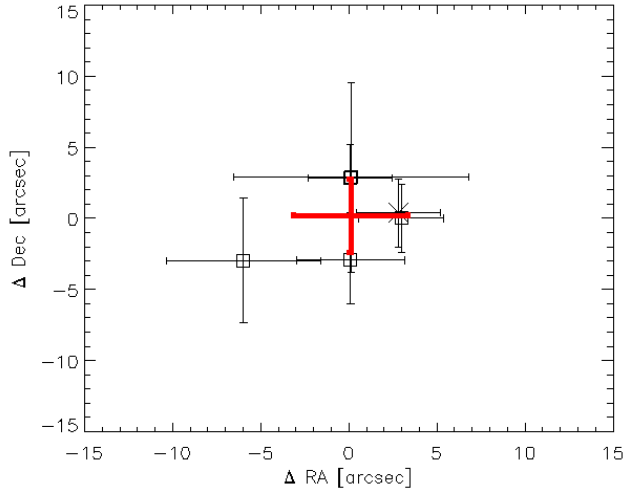} 
\caption{Measured positional offsets determined from maps of  
         five high redshift radio galaxies (squares) and by stacking 
         radio-detected AzTEC sources in the GOODS-S field ($\times$). 
         Error bars represent the 1$\sigma$ uncertainty in the 
         centroid of the 1.1~mm flux of the radio galaxies.  The red 
         thick cross denotes the mean offset of the fields with bar 
         lengths equal to the square root of the weighted 
         variance of the set of radio galaxy offsets.} 
\label{fig:radio_offsets}  
\end{figure}

To identify the correct optical/IR counterpart to MMJ065837-5557.0 we
next check for any systematic offset in the pointing, which could be
mechanical in nature or could be due to environmental influences on
the telescope at the time of the observation.  In either case,
systematic offsets are likely to vary from field to field.  Without a
known mm-bright object in our Bullet Cluster map to use as a pointing
reference (or a sufficient number of fainter objects spread
throughout the field to allow a stacking analysis) there is no way to
determine any astrometric offset specific to this field.  Instead, we
estimate our overall astrometric uncertainty as the variance in
residual pointing offsets (after the corrections applied from our
observations of PKS 0537-441) measured from similar maps of 5 high
redshift radio galaxies, along with the offsets measured from the
stacking analysis from the GOODS-S field.
Figure~\ref{fig:radio_offsets} shows the measured offsets in right
ascension and declination (physical coordinates) from these data.  The
thick red cross at $\Delta$RA, $\Delta$Dec$ =
\left(0.10\arcsec,0.18\arcsec\right)$ shows the mean offset and 
1$\sigma$ errors (3.24\arcsec\ in RA and 2.61\arcsec\ in Dec), i.e.,
consistent with no systematic specific offset. Hence, in the analysis
and discussion that follows, we apply no offset correction to the
derived centroid of the point source. We assume that the standard
deviation of systematic pointing offsets measured in other AzTEC
maps, taken in a similar manner to the Bullet Cluster data, is
representative of any positional offsets.
 
To get the most accurate centroid position for MMJ065837-5557.0, we
fit a 2-dimensional gaussian to the unfiltered image of the source.
In addition to the source position we fit for an amplitude, the
semi-major and semi-minor axes and position angle of the ellipse, and
an arbitrary DC offset.  Fitted parameters and
their statistical errors are given in Table~\ref{table:source
params}. 
 
\begin{table} 
\small 
\caption{Derived parameters for MMJ065837-5557.0.  The absolute  
         calibration-error of the flux-density is given in 
         parentheses.} 
\vspace{-4pt} 
\begin{center} 
\begin{tabular}{|l|l|} 
\hline 
Flux density  & $15.9 \pm 0.5 ~(\pm 1.3)$  mJy \\ 
SZE-corrected flux density  & $13.5 ~\pm 0.5 ~(\pm 1.0$) mJy \\ 
RA centroid position        & 06:58:37.31 $\pm 0.02\rm{s}$ \\ 
DEC centroid position       & $-55$:57:01.5 $\pm 0.32\arcsec$\\ 
Source FWHM (RA)            & $36\pm1.3$ arcseconds         \\ 
Source FWHM (Dec)           & $32\pm1.2$ arcseconds         \\ 
Position angle of elongation & $34\pm8$ degrees             \\ 
reduced $\chi^2/$ of fit    & 0.94                          \\   
\hline 
\end{tabular} 
\vspace{-6pt} 
\end{center} 
\label{table:source params} 
\end{table}

\subsection{Calibration and Flux Determination} 
 
{The AzTEC} Bullet Cluster data are calibrated using Uranus as a primary 
calibrator and adopting the technique described in \citet{wilson2008}. 
Beammaps of Uranus were made twice each night.  A linearly 
interpolated calibration factor is applied to Bullet Cluster 
observations taken between Uranus beammaps.  The calibration factor 
derived from the nearest Uranus beammap in time is used for Bullet 
Cluster observations taken before the first Uranus beammap of the 
evening and for Bullet Cluster observations taken following the last 
Uranus beammap of the evening.  Since most of the science observations 
took place following the last Uranus beammap of the evening, we 
estimate an upper limit to our overall calibration uncertainty from 
the standard deviation in reported fluxes from 31 pointing 
observations of PKS 0537-441, which is 6\%.  PKS 0537-411 is known to 
be a strongly varying source~\citep{romero1994} and so this upper limit 
should be considered robust.  Adding this in quadrature with the 5\% 
uncertainty in the brightness temperature of Uranus at 1.1~mm 
\citep{griffin1993} gives a total calibration error of 8\%. 
 
We use the 2-dimensional gaussian fit described above to estimate the flux
density of MMJ065837-5557.0 in the absence of any foreground SZE
emission from the cluster.  Since the AzTEC map has been Wiener filtered
to optimally identify point sources, an unknown degree of attenuation has been
applied to the extended SZE signal. Thus the fitted point-source flux-density,
allowing for a lower-limit on the contribution of SZE, is $S_{\rm{1.1
mm}}=13.5\pm0.5 (\pm1.0$)~mJy where the first error is the statistical error
and the second (in parentheses) is the error in absolute calibration.
 
The fitted size of MMJ065837-5557.0 is larger and slightly
elongated compared to the beam size measured by performing the same
fit to the 31 observations of the point-source PKS0537-441 (FWHM
$=29\pm1.2$ arcseconds), suggesting that the flux of MMJ065837-5557.0
may be due to a close pair, or possibly triplet (see \S 3.5), of
confused sources.  Until we have performed a more detailed subtraction
of the SZE emission, or undertaken higher-resolution observations, we
cannot rule out that the flux from MMJ065837-5557.0 is from a single
point-source.
 
\begin{figure} 
\includegraphics[width=8.0cm]{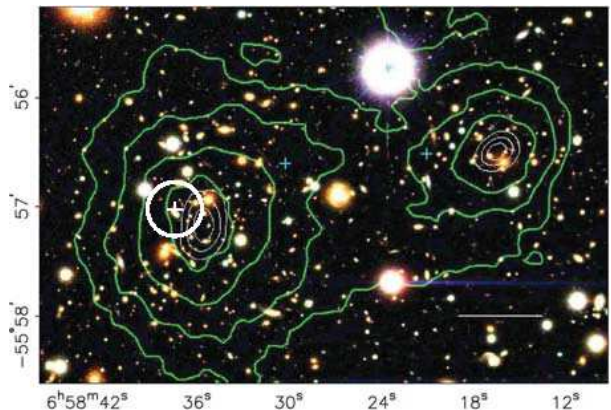} 
\caption{The weak-lensing map (green contours), from 
\citet{clowe2006} of the Bullet Cluster overlaid on an optical image 
that shows the two merging clusters of galaxies, with the less-massive
``bullet'' sub-cluster to the west of the main cluster. The white
contours towards the peaks of the mass distribution indicate their
positional uncertainty.  The position of the bright AzTEC source
MMJ065837-5557.0, which lies close to the peak of the more massive
cluster, is shown as a white cross.  The white circle indicates the
scale of the AzTEC beam with 30\arcsec\ diameter FWHM. The full AzTEC
map is significantly larger than the area of the optical image shown
here.}
\label{fig:massmap}  
\end{figure}

\section{Discussion} 
\label{sec:implications}  
 
A comparison with the measured surface-density of SMGs detected
towards wide-area blank-fields (e.g. \citet{scott2008};
\citet{coppin2006} and references therein) is sufficient to
demonstrate how unexpected is the detection of such a bright source.
Although the submillimeter source-counts are unconstrained at these
brightness-levels, a wide-area survey of $\gg 1$ sq. degree is
necessary before such an intrinsically-bright object could be expected
to be discovered in a random survey.  Two other similarly bright
sources have been found in previous SCUBA surveys: J02399-0136, with
an observed 850$\mu$m flux-density of $26\pm3$ mJy due to the
amplification by the massive cluster Abell 370 at $z = 0.37$
\citep{ivison1998}; and the high-redshift radio galaxy
8C1909+722 which shows extended submillimetre emission and has an
integrated flux of $34.9\pm3$ mJy at 850$\mu$m
\citep{stevens2003}.
 
Given that MMJ065837-5557.0 has been detected near the peak in the
weak-lensing map associated with the more massive of the two merging
clusters (see Figure~\ref{fig:massmap}), and that the Bullet Cluster
shows optical signatures (i.e. arcs) of strong lensing, magnification
of this source and amplification of the (sub)millimeter fluxes is a
natural explanation of the high apparent brightness.  Due to the low
spatial resolution of these single-dish ASTE observations (30\arcsec\
FWHM at 1.1~mm) and the astrometric uncertainty, however, we must
consider and dismiss other alternatives for the source of the
millimeter emission including Galactic stars, potential confusion with
foreground Galactic cirrus, or galaxies intrinsic to the cluster at
$z=0.297$.

\begin{figure*} 
\includegraphics[width=15cm]{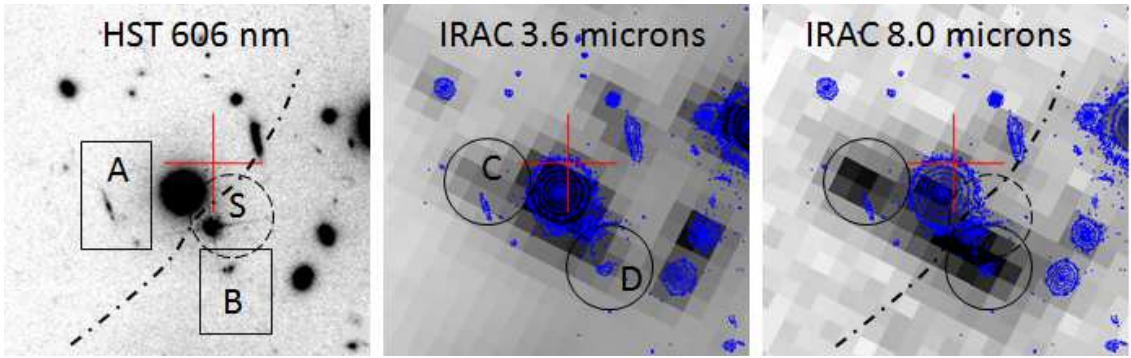} 
\caption{Left panel: HST ACS image (606W filter) showing  
a field of $23\arcsec \times 23\arcsec$ centered on the centroid 
position (shown as a red cross) of MMJ065837-5557.0.  The dimension of 
the cross indicates the $\pm 1\sigma$ positional uncertainty of 3 
arcsecs for MMJ065837-5557.0, which corresponds to a physical distance 
of $\sim 22$~kpc for a galaxy at $2 < z < 3$.  The dashed-lines, shown 
only on the HST and IRAC 8.0$\mu$m images, bisecting the two red 
galaxy systems (A/C \& B/D), and close to the position of the 
elliptical galaxy (E - a spectroscopically-confirmed member of the 
Bullet Cluster), are the approximate locations of the critical-line 
of infinite magnification for a galaxy at $z = 2.7$ (adapted from 
Gonzalez  et al. 2008). A stellar-object, S, is also identified. The 
optical galaxies A and B, enclosed within rectangles, are two 
candidates for highly-lensed background sources that may be high-$z$ 
counterparts to the AzTEC source MMJ065837-5557.0.  The IRAC 3.6$\mu$m 
(middle panel) and 8.0$\mu$m (right panel) images of the same field as 
the HST image are shown with optical contours overlaid. Clear color 
differences and spatial displacements, given in Table~1, are seen 
between the optical components A and B, and the IR components C and D 
(shown as circles). 
\citet{bradac2006} present an alternative mass model in which the 
critical-line passes closer to object A.  It requires 
higher-resolution (sub)millimeter observations with ALMA to confirm 
which of these components (if any) is the counterpart to 
MMJ065837-5557.0. Optical-IR spectroscopy are required to show whether the 
galaxy systems A/C, and B/D are multiple lensed-images of the same 
background galaxy with heavy patchy dust-obscuration, or if components 
A,B,C and D are separate galaxies involved in strong-interactions that 
stimulate an episode of luminous starformation, or alternatively if 
they are physically-unconnected galaxies in our line-of-sight.} 
\label{fig:closeup}  
\end{figure*}

\subsection{Rejecting a Galactic origin for MMJ065837-5557.0} 
 
Within a 6.5\arcsec\ radius 2$\sigma$ positional error-circle of the
AzTEC source there is a compact point-source at the resolution of HST
(object S in Figure~\ref{fig:closeup}). The object is identified as a
star in the Naval Observatory Merged Astrometric Dataset (NOMAD).  S
has an $R$-band magnitude of 18.8, which corresponds to a flux-density
of 0.19 mJy.  Assuming a blackbody photospheric spectrum ($T_{\rm eff}
\sim 4500$K) and that there is no extended circumstellar dust
envelope, which is consistent with the declining flux-density of
object S from 3.6 to 8~$\mu$m in the corresponding Spitzer imaging
(see Fig.~3), then S cannot be responsible for the 13.5~mJy emission
at 1.1~mm.  For completeness, we considered the possibility that the
"stellar-object" S is an uncatalogued quasar at an unknown
redshift. In the absence of an optical spectrum, we derive the IR
colors of source S: $S_{8.0}/S_{4.5} < 0.40$,
$S_{5.8}/S_{3.6}\sim1.7$.  Source S is bluer than a typical IR AGN and
falls outside of the region of the color-color diagram shared by SMGs
and AGN \cite{yun2008}.  We therefore reject source S as the source of
the millimeter flux.

At a Galactic latitude of $b \sim -21^o$ the foreground dust emission 
associated with Galactic cirrus is moderately-weak in the vicinity of 
the Bullet Cluster, with a 100$\mu$m surface brightness of $\sim 
5$~MJy/sr.  Adopting the measured 60/100$\mu$m cirrus color-temperature  
of $\sim 20\rm K$ (and a dust emissivity-index $\beta =2$), 
we estimate that the contribution to the measured AzTEC 1.1mm 
flux-density is $\sim 1$~mJy in the 30\arcsec\ FWHM beam. This 
estimate is sensitive to the exact temperature of the cirrus, and can 
increase to 5~mJy for colder cirrus ($T=15$K). However, the 
spatial-filter applied during the AzTEC data-reduction suppresses any 
extended foreground-emission on scales much larger than the point 
source response and therefore acts as a local sky-subtraction at the 
position of MMJ065837-5557.0.  Unless the measured power due to cirrus 
is dominated by emission on angular-scales smaller than 30\arcsec, 
foreground Galactic cirrus is unlikely to be responsible for the 
bright point-source emission of MMJ065837-5557.0. For the reasons given 
above we confidently reject a Galactic origin for MMJ065837-5557.0.

\subsection{Rejecting a cluster-member origin for MMJ065837-5557.0} 
 
The optical galaxy nearest to the AzTEC centroid is an elliptical 
galaxy, which is a spectroscopically confirmed member of the Bullet 
Cluster.  Located only 1.5\arcsec\ from the AzTEC source 
location, this $B=21.58$ elliptical galaxy at $z=0.2958$ 
\citep{barrena2002} is a natural candidate for a counterpart to 
the 1.1mm continuum emission.  It is also the nearest and the 
brightest IRAC source located within the $1\sigma$ positional 
uncertainty of MMJ065837-5557.0. 
 
\begin{figure} 
\includegraphics[width=\hsize]{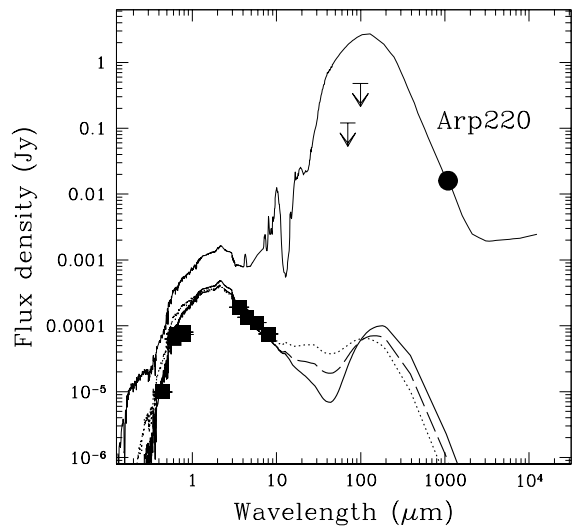} 
\caption{Spectral Energy Distribution of the $z=0.297$ cluster  
         member galaxy E (Fig.\ref{fig:closeup}) near the MMJ065837-5557.0
         position centroid.  The measured photometry data for the galaxy E are
         shown as filled squares, and they are compared with three elliptical
         galaxy SED templates by \citet{polletta2007} with different ages (2,
         5, \& 13 Gyr old shown as dotted, dashed and solid-lines
         respectively).  The two $3\sigma$ upper limits from the IRAS
         60~$\mu$m and 100~$\mu$m bands are also shown.  The Arp~220 SED
         by Polletta et al. is normalized to the 1.1mm AzTEC measurement
         (filled circle) for comparison.}
\label{fig:e0_sed}  
\end{figure}

As shown in Figure~\ref{fig:e0_sed}, both the optical and the IRAC
near-IR data of the galaxy are completely consistent with the spectral
energy distribution (SED) of a typical elliptical galaxy, dominated by
a predominantly old stellar population.  The predicted 1.1mm flux
density for such an elliptical galaxy, however, is nearly 4 orders of
magnitudes smaller than that measured by AzTEC.  However, major
mergers of two gas-rich galaxies that lead to the ULIRG phenomenon are
also known to produce an elliptical-like stellar remnant
\citep{hibbard1999}.  We explore the possibility that this elliptical
galaxy may host a substantial and compact reservoir of cold gas and
dust by plotting the SED of Arp~220 (a prototypical ULIRG) scaled to
match the measured 1.1mm flux density of MMJ065837-5557.0.  The
discrepancy between the observed flux density and the scaled Arp~220
SED is only a factor of a few in the optical bands, but the difference
grows quickly and substantially at $\lambda > 1$~$\mu$m.  The absence
of the PAH feature in the IRAC 8~$\mu$m band can be interpreted as
evidence that the elliptical galaxy does not contain a large reservoir
of cold gas and dust.  This galaxy is not detected by the Infrared
Astronomy Satellite (IRAS), and we show the $3\sigma$ upper limits of
120 mJy and 480 mJy in the IRAS 60 and 100~$\mu$m
bands\footnote{obtained using ADDSCAN/SCANPI process available through
NASA/IPAC Infrared Science Archive (http://irsa.ipac.caltech.edu).} in
Figure~\ref{fig:e0_sed}.  An Arp~220-like SED that can account for the
measured 1.1mm flux density can be ruled out with $>10\sigma$
significance.  Therefore, we confidently reject the possibility that
the elliptical galaxy is the origin of the extreme 1.1~mm continuum
flux density of MMJ065837-5557.0.

We can not confidently rule out other, more exotic alternative
possibilities as the source of the bright millimeter emission such as
nearby cooling-flow galaxies or exceptionally strong line emitters.
As described below we instead focus on the most plausible explanation:
two red IRAC sources (identified by \citet{bradac2006} and further
studied by \citet{gonzalez2008}) representing a background galaxy or
galaxies magnified by the Bullet cluster lens.

\subsection{MMJ065837-5557.0 - a highly-magnified, high-redshift starburst galaxy}  
 
The Bullet Cluster is a high-velocity merger ($\sim 4500 \rm 
kms^{-1}$) of two massive systems (the main-cluster and  
the ``bullet'' sub-cluster)  
of galaxies taking place 
in the plane of the sky.  Although there is continued 
discussion regarding the exact mass ratio ($\rm M_{main} / M_{bullet}$) 
of the two components  
ranging from 3:1 \citep{nusser2008} to 10:1 \citep{barrena2002},  
the total dark-matter  
halo cluster mass is constrained to the range 
$\sim 0.8-2 \times 10^{15} \rm M_{\odot}$ 
\citep{vikhlinin2006, clowe2006}.  MMJ065837-5557.0 is within  
15\arcsec\ of a peak in the underlying mass-distribution of the main 
galaxy-cluster and so a plausible scenario is that we are detecting 
amplified emission of a background source close to this position. 
Furthermore, the lensing models of \citet{mehlert2001} and 
\citet{gonzalez2008} show that  
critical-lines (caustics) of infinite magnification for galaxies at 
$z=3.24$ and $z=2.7$ respectively  pass within a few arcsecs of 
the centroid-position of MMJ065837-5557.0 (Fig.\ref{fig:closeup}).
Thus, strong magnification of a background submillimeter 
galaxy is our preferred explanation for the unusual brightness  
of MMJ065837-5557.0. 
We now turn to the archival HST and IRAC data to identify the most 
likely counterparts of the mm-wavelength emission. 
 
Figure~\ref{fig:closeup} shows a close-up view of the HST and IRAC fields near
MMJ065837-5557.0, along with the approximate position of the lensing
caustic-line of infinite magnification.  Having ruled out the elliptical
galaxy closest to the AzTEC centroid as the source of the strong millimeter
wavelength emission, we focus on the two reddest IRAC sources (labeled C and D
in the IRAC 8$~\mu$m image of Fig.\ref{fig:closeup}) that lie within the
uncertainty of the AzTEC centroid position, and that are also bisected by the
caustic-line.  The IRAC colors of objects C and D 
(see Table~\ref{table:photo_optIR}) are similar to those of dusty,
optically-obscured, high-redshift SMGs.
Objects C and D also appear to be the same two galaxies 
identified by \citet{bradac2006} as candidates for galaxies at $z > 6$ 
on the basis of their IR colors. We argue in 
\S~\ref{sec:photoz} that objects C and D lie at significantly lower-redshifts,
$z \sim 2.7$, a conclusion independently reached by \citet{gonzalez2008}.
 
A comparison of the HST image at 606~nm and Spitzer images at 3.6 and
8.0~$\mu$m shows that two optical sources of different morphologies
(objects A and B in Figure~\ref{fig:closeup}) lie $\le 2$\arcsec\ to
the south of the IR centroids of objects C and D, which are most
clearly defined in the 8.0$\mu$m image. A small positional offset of
RA=$-0.4$\,arcsecs, Dec=0.0\,arcsecs was applied to the archival HST
data to align it with the IRAC image. The error in this offset,
determined from the registration of stars detected in both fields, is
insufficient to account for the apparent separation of the IR and
optical pairs of galaxies.  Before considering further the possibility
of the association of the IR sources C and D with the optical HST
sources A and B, and collectively all four sources with
MMJ065837-5557.0, we present an estimate of the photometric-redshifts
of the two IR sources.

\subsection{IR photometry, colors, and photometric-redshifts} 
\label{sec:photoz} 
 
The difference in resolution between optical HST and IR Spitzer
observations is significant. While the two extended red Spitzer
sources (C \& D) are clearly seen in the original 8~$\mu$m IRAC
images, the halo of the elliptical galaxy (object E in
Figure~\ref{fig:closeup}) makes it non-trivial to measure their IR
colors. Furthermore it is difficult to unambiguously determine if the
IR emission of source C is associated with the optical emission of A,
and similarly the IR source D with that of the optical source B.  To
minimize the contamination of the elliptical galaxy on the extracted
fluxes of sources A, B, C and D, we model the elliptical galaxy using
a 2-dimensional light-profile fit to the high-resolution HST image at
606~nm. A subtraction of this model from the corresponding 775~nm data
did not reveal any significant residuals at the position of the
elliptical.  This same model was then smoothed and scaled to the
resolution and peak intensity of the elliptical galaxy emission in all
4 Spitzer IRAC bands and subtracted from each IRAC image,
producing smooth residuals at the center of the elliptical that
matched the surrounding sky values.  The optical fluxes at 606~nm and
775~nm for sources A and B were obtained using 1.5 and a 1.0~arcsec
radius aperture photometry, respectively, and adopting sky values
estimated from concentric 0.5~arcsec-width rings, after masking any
prominent objects, like star S. The IR fluxes for sources C and D at
3.6, 4.5, 5.8 and 8.0$\mu$m were extracted via aperture photometry
after subtraction of the elliptical galaxy, as explained above.  Star
S was also modeled by a PSF, scaled and subtracted from the images to
produce smooth profiles of the underlying object D, without creating
large residuals in its structure.  Sky values were estimated from
areas without contamination of faint objects within concentric 3-8
pixel rings multiple times, and 2.5--3.0 pixel (3.0--3.6~arcsec)
radial aperture photometry was performed for each estimated sky value,
and corrected to account for flux losses using the PSF model
corrections of the IRAC Data Handbook version 3.0. No attempt was made
to formally measure the errors introduced in the photometry due to the
elliptical and star subtractions, since the limited resolution of the
IRAC images is not sufficient to constrain sub-pixel positional
accuracies.  Since both C and D lie outside the centroid of E and S,
these additional errors are estimated to be less than 10\%. All
derived fluxes are listed in Table~\ref{table:photo_optIR}.

\begin{figure} 
\includegraphics[angle=90, width=8.8cm]{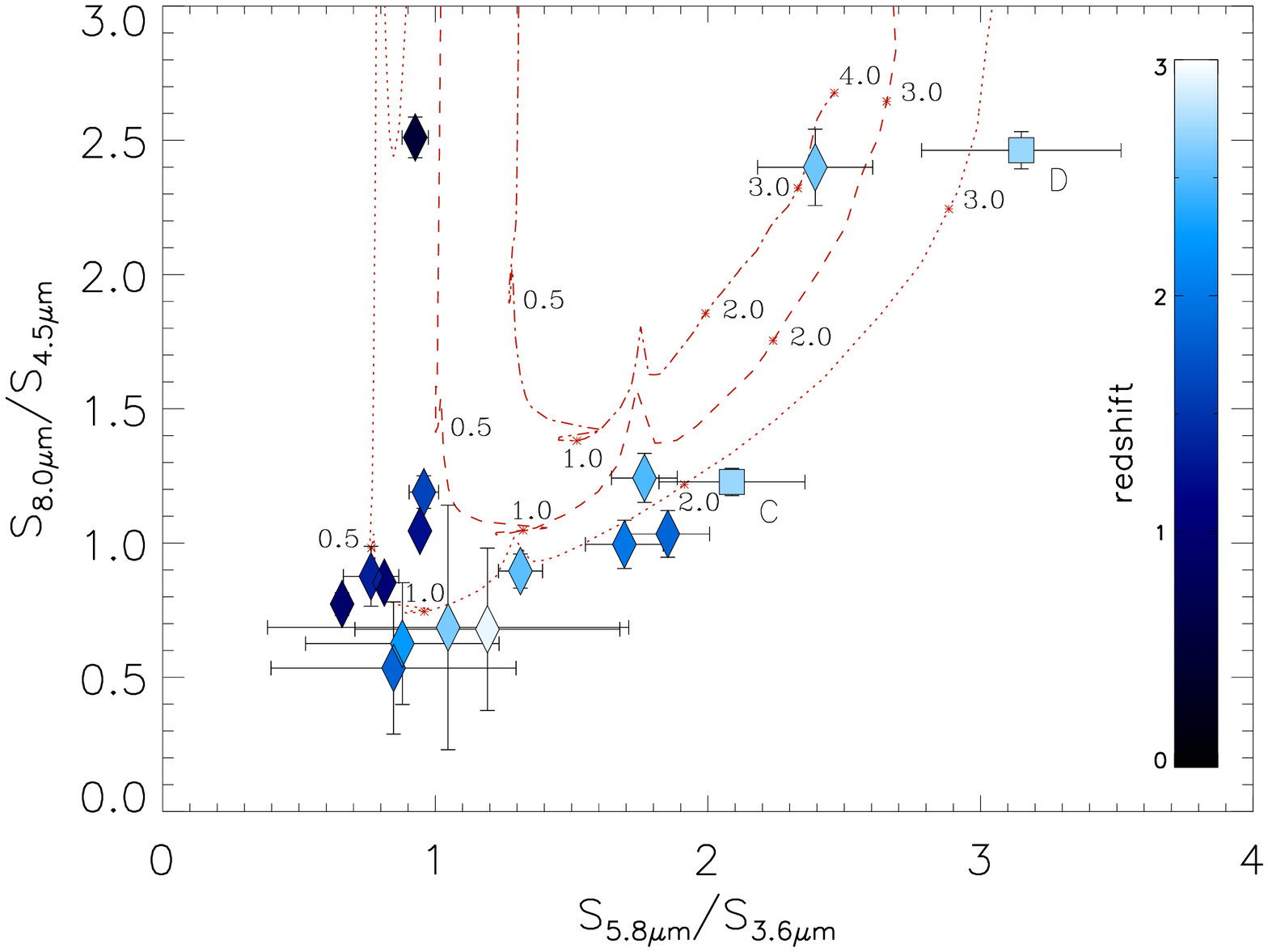} 
\caption{Color-color plot of IRAC-detected submm galaxies in the 
 GOODS-N \citep{pope2006}, SHADES SXDF \citep{clements2008} and 
 Lockman-Hole fields \citep{dye2008} with secure spectroscopic 
 redshifts, represented as diamonds, where the hue of the symbols is 
 proportional to the redshifts. SXDF850.21, at 
 $S_{8.0\mu{\rm m}}/S_{4.5\mu{\rm m}}=8.27$, $S_{5.8\mu{\rm m}}/S_{3.6\mu{\rm m}}=1.33$,  
$z_{\rm spec}=0.044$ is out of bounds of the represented plot, but has been included in the analysis. 
The squares represent the two IR bright components (C 
and D) that are possible counterparts at $z\sim 2.7$ to 
MMJ065837-5557.0.  The lines represent the color changes for  
three model spectral energy distributions of dusty galaxies 
(Siebenmorgen \& Kr\"ugel 2007) placed at increasingly larger 
redshifts, as indicated by the small labels. 
} 
\label{fig:color}  
\end{figure}

Figure~\ref{fig:color} shows an IRAC color-color 
plot, $S_{8.0\mu{\rm m}}/S_{4.5\mu{\rm m}}$ vs. $S_{5.8\mu{\rm m}}/ 
S_{3.6\mu{\rm m}}$, for 
IRAC-detected SMGs in the GOODS-N \citep{pope2006}, Subaru-XMM Deep Field 
(Clements et al. 2008), and Lockman Hole  
(Dye et al. 2008) fields, along with 
the location of sources C and D.  Both sources have IR colors 
consistent with other known SMGs. 
 
A crude photometric redshift for MMJ065837-5557.0 can be estimated,
based on the IRAC photometry, using an empirical polynomial-fit to the colors
of similar galaxies. This technique has been used successfully in the optical
and IR for various galaxy populations (e.g. \citet{connolly1995}
and \citet{pope2006}).  We adopt a functional form $z=a+\Sigma\, b_{i}\log
S_{i}$, where $a$ and $b_{i}$ are coefficients that are fit to the appropriate
galaxy population (SMGs in this case), and $S_{i}$ are the IRAC flux-densities
of the training galaxies at 3.6, 4.5, 5.8 and 8.0$\mu$m. We fit our model to
the IR colors of the 16 SMGs with unambiguous single optical/IR counterparts,
complete IRAC photometry, and robust spectroscopic redshifts in the GOODS-N
and SHADES fields (Pope et al. 2006; Dye et al. 2008; Clements et al. 2008). A
best fit results in the relationship:
 
\begin{equation}\label{eq:empfit} 
z = 1.6393 + 3.9134\log(S_{5.8}/S_{3.6}) - 2.3490 \log(S_{8.0}/S_{4.5}) 
\end{equation} 
 
\begin{figure} 
\includegraphics[width=\hsize]{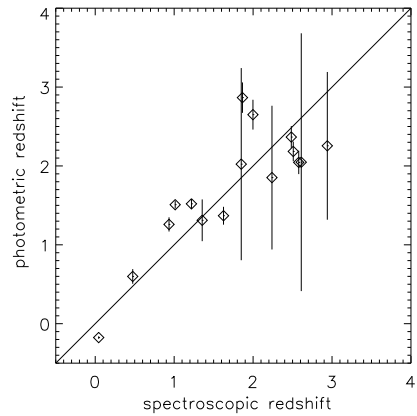} 
\caption{Photometric vs. spectroscopic redshift of  submm galaxies in the  
        GOODS-N and SHADES fields. The photometric redshift for each 
        galaxy has been calculated using an empirical fit to the 
        colors of SMGs, following the method described in \S3.3.  The 
        dispersion of values between spectroscopic and photometric 
        redshifts is $<\delta z/(1+z)> = 0.15$.} 
\label{fig:colorz}  
\end{figure}

In order to test the accuracy of this photometric-redshift estimate 
for the SMG population we use a {\em leave-one-out cross-validation} 
technique, where each member of the training-galaxy set is 
systematically excluded from the fit, and then that best-fit solution 
is used to derive its photometric redshift. This ensures that the 
accuracy of this photometric redshift technique is determined 
independently of the galaxies used to derive it. 
Figure~\ref{fig:colorz} shows the comparison of the spectroscopic 
redshifts of the training set to the photometric redshifts derived 
from the IRAC colors.  We estimate the error in each photometric 
redshift by bootstrapping on the photometric errors in the colors. We 
find an average error between photometric and spectroscopic redshifts 
of $<\delta z/(1+z)> = 0.15$ for this technique.  Although the 
accuracy of this method decreases as the redshift increases, at least 
based on this small sample, there is no apparent bias to it. 
 
Using this technique we find that both IR sources, C and D, have 
photometric redshifts of $2.7 \pm 0.2$ where the uncertainty is 
derived from the observed colours of the galaxies and 
bootstrapping on their individual errorbars.  This supports the arguments  
that they are either multiply-lensed images of the same source, or 
that they are two galaxies separated (in the image-plane) by $\sim 
35$~kpc and are undergoing an interaction that stimulates star-formation  
in dust-obscured regions.

\begin{table*}
 \begin{minipage}{170mm}
\begin{center}
\caption{Optical and IR photometry of galaxies within the positional 
         uncertainty of MMJ0659-557. Sources A--E refer to those
         objects identified in Figure~\ref{fig:closeup}. All fluxes
         are given in $\mu$Jy. Redshifts with superscripts $p$ denote
         photometric-redshift estimates (this paper), and with $s$
         denotes a spectroscopic-redshift from Barrena et al. (2002).}
\label{table:photo_optIR}
\begin{tabular}{llccccccl}
\hline
  source & redshift & 606nm & 775nm & 3.6$\mu$m & 4.5$\mu$m &
  5.8$\mu$m & 8.0$\mu$m & notes \\
\hline
A & --  & $0.8\pm0.05$ & $1.0\pm0.03$ & & & & & \\ 
B & --  & $0.6\pm0.05$ & $1.3\pm0.03$ & & & & & \\
C & $2.7 \pm 0.2^{p}$ &               & & $27\pm1$ & $44\pm1$ & $57\pm7$ & $54\pm2$ & after E subtraction\\
D & $2.7 \pm 0.2^{p}$ & & & $25\pm2$& $41\pm1$ & $80\pm8$ & $100\pm2$ &
after E subtraction\\ 
E & $0.297^{s}$ & $63\pm4$ & $80\pm4$ & $192\pm12$&
$135\pm9$& $112\pm5$ & $75\pm4$ & \\
\hline
\end{tabular}
\end{center}
\end{minipage}
\end{table*}

\subsection{\label{ref:counterparts}Counterparts to MMJ065837-5557.0} 
 
We now explore a number of scenarios regarding the possible 
associations of the optical and IR sources A, B, C and D  
(Fig. \ref{fig:closeup}). 
 
{\bf Scenario 1: Gradients of patchy dust obscuration (A$=$C and
B$=$D)}.  If we assume that the optical HST and IR Spitzer emission of
components A and C, and similarly for components B and D, arises from
different and distinct regions of the same galaxy, and that the
galaxies A/C and B/D are both at redshifts $z \sim 2.7$
(\S{\ref{sec:photoz}), then the physical separations of the IR and
optically-emitting regions in each galaxy are {\bf $< 16$~kpc} in the
plane of the lens.  We consider how this apparent separation decreases
in the source plane. Using previously published mass-models derived
from strong lensing data alone \citep{mehlert2001} along with combined
strong and weak-lensing reconstruction for the cluster 1E0657-558
\citep{bradac2006}, the image plane is mapped back to the source 
plane using the inversion software LENSTOOL \citep{jullo2007}.  We model the
bullet cluster as two large-scale smooth potentials, and include the
elliptical galaxy, a confirmed cluster-member, as an additional small-scale
local perturber.  Following \citet{bradac2006}, the main cluster is modelled
with a surface mass-density (or covergence) $\kappa(r)\,\propto \,r^{-1.2} $
and velocity dispersion $\sigma\,\sim\,1400\, {\rm kms^ {-1}} $, whilst the
sub-cluster has a slightly shallower slope of $\kappa (r) \propto r^{-0.9}$
and $\sigma\,\sim\,1200\,{\rm kms^{-1}}$.  Typically the local amplification,
$\kappa^2 - \gamma^2 $, where $\gamma$ is the shear, provides an comparable
expansion of the image-plane compared to the source-plane.  In the vicinity of
the MMJ065837-5557.0 however the shear-field of the cluster is perpendicular
to the elongation of the cluster. Hence, adopting the above composite
mass-model, the $\sim 2$arcsec separation of the pairs of optical and IR
galaxies (A/C and B/D) in the image-plane is reduced only by a factor of 0.5 -
0.75, or a separation of $1.0-1.5$ arcsec ($10-13$~kpc), in the source-plane.

It is therefore possible that the galaxies identified in the HST image
are the same as those detected by Spitzer IRAC, {\it i.e.} A=C and
B=D, implying that the thermal emission at millimeter wavelengths
arises from distant, dusty starburst galaxies with a gradient of
heavy, but patchy obscuration.  In support of this argument
recent K-band imaging of the same composite systems by
\citet{gonzalez2008} shows that the centroid of the 2.2$\mu$m emission
lies between the optical HST and longer-wavelength 3-8$\mu$m Spitzer
data.

{\bf Scenario 2: Interacting pairs of red \& blue galaxies (A$\ne$C
and B$\ne$D.)} Alternatively the HST and Spitzer data may be
indicating that we have two close pairs of strongly-interacting or
merging-galaxies, in which each pair contains an optical galaxy and an
IR dust-obscured starburst. The observed separation of sources A and C
(and equivalently B and D), {\it i.e.}  $\sim 8$ arcsecs, corresponds
to $\le 30$~kpc in source plane.  The association of composite
(possibly interacting) systems, red galaxies and blue companions, with
the environments of SMGs is not uncommon \citep{Ivison2002}.  The {\em
Antennae} galaxy provides an excellent local example of the complex wavelength-dependent morphology in a merging-galaxy with
spatially-distinct regions of luminous optical emission from young
stars, and even greater luminosity at millimeter wavelengths
originating from a young starburst population that remains
heavily-embedded in the dusty, gas-rich ISM (see for example
\citet{whitmore1999,xu2000,wang2004}).

{\bf Scenario 3: Two multiple lensed-images of a background galaxy
(A$=$B$=$C$=$D)}.  The final scenario that we consider, which is
consistent with the cluster lens model, is one in which the pairs of
optical and IR galaxies are bisected by a critical-line of infinite
magnification. In this case objects A, B, C, and D may all be multiple
lensed-images of the same background source.  Similar examples are
presented by
\citet{kneib2004} and \citet{borys2004} in the clusters Abell 2218 
($z=0.17$) and MS0451.6-0305 ($z =0.55$) respectively.  In these cases 
bright submillimeter sources detected by SCUBA (9--17~mJy at 
850$\mu$m) are associated with multiple images of highly-magnified, 
multi-component optical and IR galaxies which are bisected by the 
critical-lines. After demagnification, the intrinsic fluxes of these 
particular SMGs are 0.4--0.8~mJy at 850$\mu$m, making them some of the 
faintest SMGs detected to date. 
 
An argument against the optical sources A and B being multiple images 
of the same background galaxy is that they have different morphologies 
in the high-resolution HST observations (Figure~\ref{fig:closeup}). 
Source A shows more extended and elongated emission than source B,  
which appears to consist of two compact emitting-areas. However, the 
presence of the cluster-member elliptical galaxy (source E in 
Fig.\ref{fig:closeup}) is able to provide sufficient gravitational 
mass to distort the resolved optical images and create additional 
magnification of the background source (e.g. Natarajan \& Kneib 1997).  
It is not necessary therefore that the two lensed images of the background
source should have identical magnifications  and morphologies. 
Given the optical brightness of galaxies A and B, this scenario
can be easily tested via a spectroscopic-determination of their redshifts.

Finally, we mention that our gravitional-lens model also predicts a
third counterpart to the multiply-imaged background source with only a
modest magnification factor of 3--5 at 06h 58m 33.5s, -55d 57m 29 s
(J2000). A faint red IRAC source is seen at this position with no
optical counterpart. We do not discuss the possibility of this third
source further, except to mention that it is marginally consistent
with the orientation of the slightly-extended profile of the
millimeter emission from MMJ065837-5557.0.

Regardless of the degree of association between the IR and optical 
sources, the IRAC photometric-redshifts are well-within the redshift 
distribution of the blank-field SMG population, and thus we believe 
that the most plausible explanation for the mm-wavelength flux is that 
1) the AzTEC source MMJ065837-5557.0 lies behind the Bullet Cluster, 2) it 
is lensed by the Bullet Cluster, and 3) it is a product of the 
combined emission of the IRAC objects C and D.  This scenario also 
naturally explains the approximate equal distance from the AzTEC 
centroid position to the two IR galaxies. In the absence of a detailed 
lensing model in the vicinity of the AzTEC source and/or spectroscopic 
confirmation, we assume in the remainder of this paper that both pairs 
of optical/IR sources (A/C and B/D) are associated with MMJ065837-5557.0,  
which is a lensed background SMG.

\subsection{Contribution of the Sunyaev-Zel'dovich effect to the observed  
flux of MMJ065837-5557.0} 
 
Two classes of effects can serve to increase the measured brightness
of detected point sources: multiplicative effects (lensing) and
additive contributions to the flux due to mm-wavelength emission along
the line of sight to the source.  MMJ065837-5557.0 clearly suffers
from both.  In this section we describe the non-negligible
millimeter-wavelength emission due to the thermal Sunyaev-Zel'dovich
Effect (SZE, \citet{SunZel1970, SunZel1972}) that is added to the
intrinsic lensed emission of MMJ065837-5557.0. The thermal SZE is
produced by inverse-Compton scattering of the CMB photons off
electrons in the hot and dense inter-cluster medium (ICM) of
clusters. We do not consider the contribution from the weaker kinetic
SZE that is due to the bulk motion (velocity) of the cluster plasma
with respect to the CMB reference-frame. Consequently the thermal SZE
is a dominant source of secondary-fluctuations on the surface
brightness of the CMB on small angular scales, less than a few
arcminutes.

Unable to perform a multi-wavelength spectral decomposition of the SZE
and the bright background galaxy MMJ065837-5557.0, with AzTEC 
observations only at 1.1mm, we can estimate the level of the thermal 
SZE contamination in two independent ways. 

First, taking advantage of the spatial-resolution of the AzTEC
observations and assuming that the SZE emission from the cluster is
smooth on spatial scales smaller than that of the AzTEC beam
(30\arcsec ), we can spatially-filter the AzTEC map and fit for the
1.1mm flux consistent with only point-source emission (as discussed in
\S2.1).  Alternatively, as described below, we can model the SZE
emission from the Bullet cluster with a simple model for the
electron-temperature and density distribution to estimate the 1.1mm
flux from the SZE at the position of the AzTEC point-source.

The derivation of the intensity of the SZE and its observed properties  
at radio to millimeter-wavelengths have been reviewed extensively 
\citep{SunZel1980, birkinshaw1999, carlstrom2002, birkinshaw2007}.   
Briefly, considering only the thermal distortion, 
the frequency-dependent SZE is given by the simple expression  
 
\begin{equation}  
\label{eq:kay01} 
\delta I_\nu = I_0 [g(x) y]   , 
\end{equation}  
where $I_0 = 2 (k_B T_0 )^3 / ( h c )^2 $ and $T_0 = 2.725$K is 
the mean temperature of the CMB. The Comptonization 
parameter, $y$, equal to the optical-depth times the 
fractional energy-gain of each scattering event, is  
 
\begin{equation}  
\label{eq:kay04} 
y = \frac{k_B \sigma_T} {m_e c^2} \int n_e T_e \textrm{d} l \quad. 
\end{equation}  
 
Finally the dimensionless spectral function, $g(x)$, where $x = h \nu / k_B T_0$, 
describes the frequency-dependence of the thermal SZE. Neglecting relativistic 
corrections, $g(x)$ has the form 
 
\begin{equation}  
\label{eq:kay02} 
g(x) = \frac{x^4 e^x} { (e^x -1)^2} \Bigg( x \frac{e^x + 1} {e^x - 1} 
- 4 \Bigg) \quad. 
\end{equation}

Since the intensity of SZE is dependent only on the line-of-sight
integral of the electron pressure through the Bullet cluster, we
describe below our estimations for the local value of the electron
temperature and density at the position of MMJ065837-5557.0.
 
It is common to adopt an isothermal assumption within the cluster
density distribution (typically a $\beta$-model), or at least a weak
radial-dependence on the temperature of the intracluster-medium within
the virial radius \citep{irwin1999}. Support for this assumption is
also found in hydrodynamical simulations \citep{eke1998, pearce2000}.
In the case of the Bullet cluster, \citet{markevitch2002} have
provided a temperature-map that shows a gas-temperature variation
across the merging clusters between 10--24 keV, ignoring the
concentrated cold-spot or "bullet" (with a temperature of 6--7\,keV).
The spatial resolution of AzTEC (30" FWHM beamsize) is sufficient to
estimate that the local temperature of the X-ray-emitting gas, at the
position of MMJ065837-5557.0, is 14--16 keV (regions f--g in Fig.2 of
\citet{markevitch2002}), or similar to the average cluster temperature 
of $\sim 15$\,keV.

Whilst the density-profile of the gas to the west of the main X-ray
peak (in the direction of the "bullet") requires the inclusion of a
shock-front and an additional component to represent the lower-mass
merging cluster, the situation is less-complicated to the east of the
X-ray peak brightness, in the direction of MMJ065837-5557.0.  The
local electron-density, $n_e$, in the eastern hemisphere (at the
position of MMJ065837-5557.0) is calculated using a single-component
spherically-symmetric $\beta$-model ($\beta$=0.7) centered at the
primary peak of the observed X-ray surface-brightness distribution.
We begin by considering a range for the total cluster-mass,
$M_{tot}=0.8-1.9 \times 10^{15} M_{\odot}$ \citep{clowe2006,
vikhlinin2006}.  Following \citet{kay2001} we then assume that the ICM
is fully ionized, with a helium mass fraction of $0.24$, such that
$n_e = 0.88\rho/m_H$, where $m_H$ is the mass of a hydrogen atom. The
baryonic mass density, $\rho$, is calculated assuming a global baryon
fraction $f_b = M_b / M_{tot} = 0.06 h^{-3/2}$ \citep{ettori1999}.
The exercise was then repeated assuming an alternative $\beta$-model
centered on the peak of the weak-lensing signal from the main cluster
that dominates the total cluster mass (\S.3.3).  Finally, provided
that the electron temperature, $T_e$, is similar to the X-ray gas
temperature, we determine that the contribution of the SZE at 1.1mm at
the position of the AzTEC point-source MMJ065837-5557.0 is $\sim 1.5 -
4.2$\,mJy/beam.  This estimated range of underlying flux due to the
SZE is consistent with what we measure directly from the extended
component in the spatially-resolved 1.1mm emission at the position of
the AzTEC point-source (\S 2.1).

We therefore conclude that the 1.1mm flux-density of MMJ065837-5557.0
due solely to continuum emission from a lensed high-redshift SMG is in
the range 12-14~mJy.  An analysis of the detected extended AzTEC
emission towards the Bullet Cluster (Ezawa et al. in preparation) will
provide more accurate information on the location of MMJ065837-5557.0
with respect to the peak of the SZ effect, and consequently a more
accurate measurement of the point-source contamination to the
integrated SZE (a potential problem for lower-resolution experiments)
than the estimate described above.
 
Assuming a range of spectral-indicies $\alpha \sim 3 - 4$ (where $\rm
F_{\nu} \propto \nu^\alpha$) and dust temperatures to describe the
rest-frame spectral energy distribution of MMJ065837-5557.0 in the
millimeter regime, we also determine that MMJ065837-5557.0 is too
faint to be detected in the low-resolution (FWHM$\sim4.5$\arcmin)
ACBAR observations at 1.4~mm given the published noise-level of
30$\mu$K \citep{gomez2004}. This is consistent with the lack of
comment by Gomez et al. about a detectable signal at the
null-frequency of the SZE.

}

\subsection{The intrinsic luminosity of MMJ065837-5557.0  
- a dusty high-redshift LIRG} 
 
The magnification map of the Bullet Cluster by \citet{mehlert2001} 
suggests a robust and conservative lower-limit of $A=20$ for the 
amplification of MMJ065837-5557.0.  More typical amplifications of 
$A=1-3$ have been found for SMGs behind low-z massive clusters 
\citep{Ledlow2002, knudsen2006}.  
It is the proximity of MMJ065837-5557.0 to a lensing caustic-line that
makes this such an exceptional case.  The implication for such a large
magnification is that the AzTEC source has an intrinsic flux of
$S_{\mathrm{1.1mm}}< 0.68$~mJy, after subtraction of the expected
contribution from the Sunyaev-Zel'dovich effect.  This is a factor of
$>3$ lower than the flux of the faintest blank-field SMGs detected in
the full 225 sq. arcmin AzTEC field towards the Bullet Cluster (which
occupies only the central 12 sq. arcmins).
 
Assuming that the AzTEC point-source is a background galaxy at $z \sim
2.7$, then then intrinsic millimeter-wavelength brightness of
MMJ065837-5557.0, after the corrections for the SZE contribution and
amplification, corresponds to a FIR luminosity of $< 10^{12} \rm
L_{\odot}$.  Given the relative insensitivity of the
millimeter-wavelength flux-density as a function of redshift, this is
a robust result provided that the AzTEC source lies in the redshift
range $\sim 1 < z < 8$. Hence, rather than having discovered an
extremely rare example of a hyperluminous dust-obscured galaxy,
MMJ065837-5557.0 is more typical of a luminous IR galaxy (LIRG) with
$L_{\rm{FIR}}\le 10^{11} - 10^{12} \rm L_{\odot}$ in the high-redshift
Universe, with a star-formation rate (SFR) in the range $\rm 5-50
M_{\odot}/yr^{-1}$
\citep{Kennicutt1998}.  

The conservative lower-limit on the magnification implies that even a
more modest SFR may be appropriate, closer to that of only a
mildly-active starburst galaxy.  This would support the scenario in
which the millimeter emission arises from a distant starburst with
patchy dust obscuration and a low dust-content, which in turn
increases the likelihood of detecting both the optical and IR emission
from the source (i.e. A/C and/or B/D).  Only high-resolution
(sub)millimeter interferometric imaging of MMJ065837-5557.0, to
unambiguously identify the counterpart(s), combined with spectroscopic
measurements to determine the redshifts (and possibly velocity
dispersion) of all optical and IR components (A, B, C, D) will allow
progress in understanding the nature of this potentially rare example
of a low to intermediate luminosity millimeter-selected galaxy in the
high-redshift Universe.
  
\section{Conclusions} 
\label{sec:conclusions}  
 
Sensitive continuum observations at 1.1mm with the AzTEC camera 
installed on the 10-m diameter Atacama Submillimeter Telescope 
Experiment have been made towards the Bullet cluster.  
An extremely bright point-source (MMJ065837-5557.0) is detected  
with an observed flux-density at 1.1mm of 15.9\,mJy. The centroid  
position of MMJ065837-5557.0 lies close to the largest mass-density  
peak in the weak-lensing map of \citet{clowe2006}.

Archival IR (Spitzer IRAC) images show the presence of two red
sources, separated by $\sim 8$~arcsecs (or $\le 30$\,kpc in the
source-plane), that lie within the positional-error of
MMJ065837-5557.0, and that also straddle the critical
magnification-line ({\it i.e.} a region of extreme amplifications) of
\citet{mehlert2001}. The Spitzer IRAC colors of these two sources are
typical of high-z SMGs, and photometric-redshifts place both sources
at $z\sim 2.7 \pm 0.2$.  The proximity of HST-identified optical
sources to both of these Spitzer IRAC galaxies leaves open 2
possibilities: 1) that each optical/IR pair are individual galaxies
undergoing close interaction, or 2) that each pair is in fact only one
galaxy with a complex morphology of patchy dust-obscuration.

Having rejected a Galactic origin and a cluster member origin as the
source of the AzTEC emission, the above conditions suggest that
MMJ065837-5557.0 is a background source associated with one or both of
the red IR Spitzer sources, and is strongly-lensed by this massive
cluster.  The agreement in the location of the critical-line in the
magnification maps of \citet{mehlert2001}, \citet{bradac2006} and
\citet{gonzalez2008} for this cluster place a strong conservative
limit of $> 20$ for the magnification of the millimeter emission from
MMJ065837-5557.0.  There exist many similarities between this example
of an extremely bright millimeter source, and the possibility of
gravitationally-lensed interacting galaxies that produce a bright
submillimeter arc in MS0451.6-0305 \citep{borys2004}.
 
After subtracting the contribution ($\le 25\%$) due to the extended SZE  
at 1.1mm from the total AzTEC continuum flux of MMJ065837-5557.0, we  
estimate that the intrinsic (i.e. demagnified) FIR luminosity of  
MMJ065837-5557.0 is  
$<10^{12}\rm L_{\odot}$ provided the source lies in the redshift  
range $\sim 1 < z < 8$. Hence MMJ065837-5557.0 is  
representative of the fainter population of starforming galaxies 
that is undetectable, without the assistance of lensing,   
with the current generation of small single-dish 
(sub)millimeter telescopes due to source-confusion and lack of 
sensitivity. Until the commissioning of the next-generation of larger 
millimeter-wavelength facilities (e.g. LMT, ALMA) the only 
opportunity to study this fainter luminosity class of IR galaxies,  
that dominate the production of the cosmic infrared background, is to 
take advantage of the significant amplification that exist towards the 
critical magnification-lines of massive nearby clusters.

\section*{acknowledgments} 
We thank the referee for the thorough reading and suggestions to 
improve the paper, and also Jack Hughes for useful comments. 
We recognise and thank Douglas Clowe for valuable 
discussions during the early preparation of this paper.  
This study was financially supported by 
MEXT Grant-in-Aid for Scientific Research on Priority Areas No.\ 
15071202, by the Grant-in-Aid for the Scientific Research from the 
Japan Society for the Promotion of Science (No. 19403005), by NSF 
Grant \#0540852, and CONACYT grants \#50786 and 60878.  Some of the 
data analyzed in this paper were obtained from the Multimission 
Archive at the Space Telescope Science Institute (MAST) and the 
Spitzer Space Telescope archive. STScI is operated by the Association 
of Universities for Research in Astronomy, Inc., under NASA contract 
NAS5-26555.  The Spitzer Space Telescope is operated by the Jet 
Propulsion Laboratory, California Institute of Technology, under a 
contract with NASA.  This research has made use of NASA's Astrophysics 
Data System.

\bibliography{bullet_point} 
 
\end{document}